\documentclass[pre,floatfix,twocolumn]{revtex4}

\usepackage{amsmath,amssymb,latexsym,epsfig,graphics,epsf}
\usepackage{graphics}
\usepackage{float}
\newcommand{\be}{\begin{equation}}
\newcommand{\ee}{\end{equation}}

\newcommand{\Sex}{s_{\rm ex}}
\newcommand{\sex}{s_{\rm ex}}

\usepackage{verbatim}
\usepackage{times}

\newcommand{\angleb}[1]{\left\langle #1 \right\rangle}

\newcommand{\bfa}[1]{\mathbf{#1}}

\begin{document}

\title{The variation of the dynamic susceptibility along an isochrone}
\date{\today}
\author{Nicholas P. Bailey}
\affiliation{DNRF Center ``Glass and Time'', IMFUFA, Dept. of Sciences, Roskilde University, P. O. Box 260, DK-4000 Roskilde, Denmark}
\author{Thomas B. Schr{\o}der}
\author{Jeppe C. Dyre}
\affiliation{DNRF Center ``Glass and Time'', IMFUFA, Dept. of Sciences, Roskilde University, P. O. Box 260, DK-4000 Roskilde, Denmark}


\begin{abstract}

Koperwas {\it et al.} showed in a recent paper, Phys. Rev. Lett. {\bf 111}, 
125701 (2013), that the dynamic susceptibility $\chi_4$ as estimated by dielectric measurements for certain glass-forming liquids decreases substantially 
with increasing pressure along a curve of constant relaxation time. This observation is at odds with other measures of dynamics
being invariant and seems to pose a problem for theories of glass formation. We
show that this variation is in fact consistent with 
predictions for liquids with hidden scale invariance: measures of dynamics at constant volume are invariant along isochrones, called isomorphs in such
liquids, but contributions to fluctuations from long-wavelength fluctuations can vary. This is related to the known non-invariance of the isothermal bulk modulus. Considering the version of $\chi_4$ defined for the NVT ensemble, data from simulations of a binary Lennard-Jones liquid show in fact a slight increase with increasing density. This is a true departure from the formal invariance expected for this quantity.

\end{abstract}

\maketitle

\section{Introduction}

A quantity of great interest in recent years in the context of supercooled, glass-forming liquids is the dynamic susceptibility $\chi_4$, associated with a four-point correlation function $S_4(k,t)$. This was originally introduced to understand spin-glass models \cite{Kirkpatrick/Thirumalai:1988}, subsequently used to study dynamical heterogeneity in computer simulations of glass-forming liquids \cite{Dasgupta/others:1991,Donati/others:1999b,Glotzer/Novikov/Schroder:2000,Lacevic/others:2003,Andersen:2005,Toninelli/others:2005}, and with the introduction of experimentally accessible approximations to $\chi_4$, in real liquids \cite{Berthier/others:2005,Berthier/others:2007a,Berthier/others:2007b,Dalle-Ferrier/others:2007}. It quantifies the dynamical heterogeneities; in principle it can interpreted in terms of a length scale $\xi_4$ characterizing dynamical fluctuations or the number of dynamically correlated particles during structural relaxation, but the precise relation is not trivial \cite{Flenner/Szamel:2010}; even assuming a scaling $\chi_4\sim \xi_4^3$, the proportionality factor is non-universal \cite{SzamelPrivate:2014}. The growth of $\chi_4$ (or $\xi_4$) is considered relevant for explaining the dramatic dynamical slow-down as the glass transition is approached, as well as other features of relaxation in viscous liquids \cite{Berthier/others:2011}. In particular, it is believed that there is a unique relation between the relaxation time $\tau$ and the shape of the relaxation spectrum on the one hand \cite{Ngai/others:2005}, and the size of dynamical heterogeneities \cite{Adam/Gibbs:1965,Parisi:1999, Xia/Wolynes:2000,Xia/Wolynes:2001,Biroli/Bouchaud:2004, Lubchenko/Wolynes:2007} on the other hand. As a recent example of such a claim, Flenner {\it et al.} described a universal behavior in the relation between $\xi_4$ and $\tau$ rescaled by its value when violations of the Stokes-Einstein relation become apparent \cite{Flenner/Staley/Szamel:2014}. We note here that $\chi_4$ is not uniquely defined: it depends on the correlator of interest, and, as will be discussed in detail below, on the statistical ensemble (NVE, NVT, etc.).

Pressure has been increasingly exploited  as an extra experimental parameter \cite{Roland/others:2005}. This has led to a focus on liquids obeying so-called power-law density scaling when temperature $T$ and pressure $p$ are varied; that is, liquids for which the relaxation time depends only on a scaled quantity $\Gamma\equiv \rho^\gamma/T$ where $\rho$ is the density and $\gamma$ a system-specific scaling exponent \cite{ Alba-Simionesco/Kivelson/Tarjus:2002, Dreyfus/others:2003, Alba-Simionesco/others:2004, Dreyfus/others:2004, Casalini/Roland:2004, Roland/others:2005}. An experimentally observed feature of such liquids is so-called isochronal superposition according to which the relaxation spectra corresponding to the same relaxation time -- but different densities and temperatures -- superpose \cite{Roland/Casalini/Paluch:2003, Ngai/others:2005, Mierzwa/others:2008,Capaccioli/others:2012} (see Ref.~\onlinecite{Casalini/Roland:2013} for an apparent exception). This suggests that the physics governing relaxation is the same at points in the phase diagram for which the relaxation time is the same. 

Given the existence of such liquids, and the supposed link between $\tau$ and $\chi_4$, it is natural to investigate the behavior of $\chi_4$ along isochrones.  This was recently done for the van der Waals glass-forming liquids $o$-terphenyl, glibenclamide and phenylphthalein-dimethylether in Ref.~\onlinecite{Koperwas/others:2013}. Surprisingly, a significant variation was found, with the maximum $\chi_4$ decreasing as pressure (and therefore density) increased along an isochrone (see also Refs.~\onlinecite{Grzybowski/others:2012a,Grzybowski/others:2013}); in Ref.~\cite{Alba-Simionesco/Dalle-Ferrier/Tarjus:2013} the opposite behavior was reported for dibutyl-phthalate; in Ref.~\onlinecite{Fragiadakis/Casalini/Roland:2009} no significant variation was found for four other liquids. Where a variation was seen, the interpretation was that the temperature- and density-related contributions to the dynamical heterogeneities are non-equivalent, since they contribute differently to different measures of dynamics. Thus the postulated unique relation between $\chi_4$ and the growth of $\tau$ must apparently be questioned; for instance Alba-Simionesco {\it et al.} consider the variation along an isochrone of the number of correlated particles to contradict predictions from the Random First-Order Transition theory \cite{Alba-Simionesco/Dalle-Ferrier/Tarjus:2013}. On the other hand it is generally believed that, e.g., for van der Waals liquids, density and temperature changes affect the dynamics in the same way along an isochrone \cite{Ngai/others:2005}, raising the question: Is this wrong or is the traditional $\chi_4$-quantity not the relevant measure of dynamic heterogeneities? We note that simulation results showing invariance of the NVT quantity in Lennard-Jones systems were presented a few years ago \cite{Coslovich/Roland:2009,Gnan/others:2009}; the issue of ensemble-dependence was not discussed though.

The purpose of this article is to throw light on this question using isomorph theory, which provides a theoretical framework for the density-scaling behavior mentioned above. Following extensive theoretical and simulation investigations \cite{Bailey/others:2008c,Gnan/others:2009,Ingebrigtsen/others:2012, Dyre:2013} we have proposed the existence of a class of simple liquids \cite{Ingebrigtsen/Schroder/Dyre:2012a}. We use the term Roskilde- (R) liquids to distinguish from earlier senses of liquid simplicity \footnote{In earlier publications we used the term ``strongly correlating liquids'' referring to the $NVT$-equilibrium correlations between potential energy and virial, but this was often confused with strongly correlating quantum systems.}. Their key feature is the existence of isomorphs: curves in the phase diagram along which several properties are invariant to a good approximation \cite{Gnan/others:2009}, including all dynamical quantities, as long as volume is not allowed to fluctuate. One of these properties is the relaxation time in appropriate reduced units; this suggests that liquids obeying density-scaling (not necessarily power-law density scaling) can be identified as simple liquids in the R sense, and that isomorphs in experiments can be identified with isochrones.

\section{\label{isomorphs}Isomorphs}

The theory of isomorphs---the formal theory underlying
the concept of R liquids---takes as its starting point the following
general definition of isomorphic state points: Two state points $(\rho_1, T_1)$ 
and $(\rho_2, T_2)$  are isomorphic if the Boltzmann factors of corresponding
microstates are proportional:

\begin{equation}\label{proportional_Boltzmann}
\exp\left(-\frac{U(\bfa{r}_1^{(1)}, \ldots, \bfa{r}_N^{(1)})}{k_BT_1}\right)
 = C_{12} \exp\left(-\frac{U(\bfa{r}_1^{(2)}, \ldots, \bfa{r}_N^{(2)})}{k_BT_2}\right)
\end{equation}
Here $U$ is the potential energy function and $C_{12}$ depends on the two state
points, but not on which microstates are considered. Corresponding microstates 
means $\rho_1^{1/3} \bfa{r}_i^{(1)} = \rho_2^{1/3} \bfa{r}_i^{(2)}$, or 
$\bfa{\tilde r}_i^{(1)}=\bfa{\tilde r}_i^{(2)}$, where a tilde denotes so-called 
reduced units. Reduced units for lengths means multiplying by $\rho^{1/3}$, 
for energies dividing by $k_BT$, and for times dividing by 
$(m/k_BT)^{1/2}\rho^{-1/3}$ (for Newtonian dynamics). An isomorph is a curve in the phase diagram consisting of points which are isomorphic to each other.
From the definition it follows that all structural and dynamical correlation
functions are invariant when expressed in reduced units. We speak of a quantity being {\em formally isomorph invariant} if its invariance follows from definition~\eqref{proportional_Boltzmann}; because the proportionality of Boltzmann factors is typically only approximate, the actual extent to which a given quantity is invariant has to be checked empirically. Thermodynamic quantities which do not involve volume derivatives, such as the excess entropy $S_{\textrm{ex}}$ and specific heat at constant volume $C_V$, are also formally isomorph invariant. We define the density scaling exponent 

\begin{equation}\label{gamma_definition}
\gamma\equiv(\partial\ln T/\partial\ln\rho)_{\Sex}
\end{equation}
as the slope of isomorphs in $(\ln\rho,\ln T)$ space \cite{Gnan/others:2009}. To a good approximation this depends only on temperature; this is equivalent to approximating the potential energy of a Roskilde liquid with the following form \cite{Dyre:2013}

\begin{equation}\label{quasiuniversalU}
U(\bfa{R}) \cong h(\rho) \tilde \Phi(\tilde{\bfa{R}}) + g(\rho)
\end{equation}
Here $\tilde{\bfa{R}}$ is the 3N-dimensional vector of positions in reduced units (i.e., multiplied by $\rho^{1/3}$), and $\tilde \Phi(\tilde{\bfa{R}})$ is a dimensionless function of the reduced coordinates. The scaling function $h(\rho)$ determines the shapes of isomorphs via 

\begin{equation}\label{h_rho_T_const}
h(\rho)/T=\textrm{const};
\end{equation}
its logarithmic derivative 
\begin{equation}d\ln h/d\ln \rho
\end{equation}
is just $\gamma$. The term $g(\rho)$ depends on density (or volume) but not on the microscopic coordinates \cite{Dyre:2013}: it contributes an extra non-invariant part to the free energy and its volume derivatives, including the bulk modulus. That this density-dependent term is actually non-local plays a role later on in the discussion. More recent developments of isomorph theory allow for the variation of $h(\rho)$ from one isomorph to another, or equivalently, that the exponent $\gamma$ can depend on temperature at fixed density \cite{Bailey/others:2013, Schroder/Dyre:2014}. A general method for identifying isomorphs is to consider the configurational adiabats, which are formally isomorphs (that is, the excess entropy is formally isomorph invariant); their slope in $(\ln \rho, \ln T)$ space is given by the fluctuation formula

\begin{equation}\label{gamma_fluctuations}
\gamma = \frac{\angleb{\Delta U\Delta W}}{\angleb{(\Delta U)^2}},
\end{equation}
which when combined with Eq.~\eqref{gamma_definition} allows the curves to be generated in a step-wise manner.

A crucial insight from this framework is that any quantity which is claimed to control, for example, the relaxation time must also be invariant on isomorphs. Thus it is vital to consider the isomorph invariance of different formulations of $\chi_4$---only one which is formally isomorph invariant can be relevant. We show below that the version of $\chi_4$ which has been estimated experimentally is {\em not} invariant on an isomorph, because the isothermal bulk modulus is not. We also show that an isomorph-invariant version of $\chi_4$ does exist, namely the version defined for the $NVT$ ensemble. 

\section{\label{isomorphInvarianceChi4}Isomorph invariance of $\chi_4$}

We consider the different versions of $\chi_4$ and the approximate expressions used to determine it experimentally (see Ref.~\onlinecite{Maggi/others:2012} for an experimental determination which does not use these approximations). For simplicity, in this section we consider a pure substance, so that concentration fluctuations need not be accounted for. Recall that $\chi_4$ can be defined as the variance of the correlator whose average is some correlation function of interest; thus it measures dynamical fluctuations. If we consider a two-time equilibrium correlation function $C(t)$, we can write $C(t) = \angleb{C_2(t_0,t_0+t)}$ where $C_2(t_0,t_1)$ is the fluctuating two-time correlator. Here the average may be interpreted as over initial configurations keeping the initial time $t_0$ fixed and/or, as is done in practice with simulation data, over different initial times $t_0$ within the same trajectory. One now defines 

\begin{equation}
\chi_4(t)\equiv N \sigma_{C}^2 = N \left(\angleb{C_2^2} - C(t)^2\right).
\end{equation}
where the total number of particles $N$ is included to give an intensive quantity. 

Since $\chi_4$ measures fluctuations, it should not be too surprising that it depends on ensemble. Berthier {\it et al.} \cite{Berthier/others:2007a} have analyzed the ensemble dependence in detail. When going from an ensemble with a constrained global variable to one where it is free to fluctuate,  $\chi_4(t)$ increases by a positive amount corresponding to the fluctuations 
induced by those of the unconstrained variable. For example, in going from NVE to NVT the energy is allowed to fluctuate; the additional contribution to $\chi_4$ involves the energy fluctuations themselves and the correlation between $C$ and $E$; this leads to a term involving the isochoric specific heat and the temperature derivative of $C(t)$:

\begin{equation}\label{chi4_NVT_NVE_T_term}
\chi_4^{\textrm{NVT}}(t) = \chi_4^{\textrm{NVE}}(t)
 + \frac{1}{c_V/k_B}\left(\frac{\partial C(t)}{\partial \ln T}\right)_\rho^2
\end{equation}

Going further by allowing the volume to fluctuate leads to the NPT ensemble and
an additional term involving the isothermal bulk modulus and the density derivative of $C(t)$ \cite{Berthier/others:2007a}:

\begin{align}
\chi_4^{\textrm{NPT}}(t)  &= \chi_4^{\textrm{NVT}}(t) + \frac{\rho k_BT}{K_T}\left(\frac{\partial  C(t)}{\partial \ln\rho}\right)_T^2 \\
&= \chi_4^{\textrm{NVE}}(t)
 + \frac{1}{c_V/k_B}\left(\frac{\partial C(t)}{\partial \ln T}\right)_\rho^2
 + \frac{\rho k_BT}{K_T}\left(\frac{\partial C(t)}{\partial \ln\rho}\right)_T^2 \label{chi4NPTdecomp1}
\end{align}
Here $c_V=C_V/N$ is the isochoric specific heat per particle, and $K_T$ the isothermal bulk modulus. For a pure substance as considered here this expression for $\chi_4^{\textrm{NPT}}$  is equivalent to the ensemble-independent quantity obtained by taking the $k\rightarrow$ limit of $S_4(k,t)$ after the thermodynamic limit (see Appendix~\ref{S4_definition}); for mixtures additional terms related to concentration fluctuations must be included. The above relation is based on the formalism for transforming between ensembles developed by Lebowitz {\it et al.}  \cite{Allen/Tildesley:1987,Lebowitz/Percus/Verlet:1967} but applied to the variance of a two-time dynamical quantity. In this formalism it must be realized that by the $NVT$ ensemble is meant an ensemble of constant-energy trajectories with different energies and same volume undergoing Newtonian dynamics (i.e. at constant energy, as opposed to for example Brownian dynamics which also samples the NVT ensemble), while by $NPT$ is meant an ensemble of similar Newtonian trajectories only now with volume as well as energy varying from one member of the ensemble to another. Physically this corresponds to fluctuations of energy and density for a given region being slow compared to the microscopic relaxation processes; thus individual
relaxation events in this picture are not directly affected by changes in the system's energy or volume because the latter take place on longer time scales.

Of the three terms in Eq.~\eqref{chi4NPTdecomp1}, we show now that the first two are isomorph invariant, while the third is not. The first term is a measure of fluctuations in NVE dynamics. From the definition of isomorphs it follows that the reduced unit forces for corresponding configurations from different members of the isomorph are the same, and thus that Newtonian (NVE) dynamics is isomorph invariant \cite{Gnan/others:2009}. While equivalence of individual trajectories will be spoiled by deviations from perfect invariance and their chaotic nature, it follows that statistics based on the trajectories, including $\chi_4^{\rm NVE}$, are isomorph invariant. To investigate the invariance properties of the second and third terms (for brevity, hereafter referred to as the $T$- and $\rho$-terms respectively) in Eq.~\eqref{chi4NPTdecomp1}, we first consider the derivatives of an arbitrary isomorph invariant quantity $X$. Since the excess entropy per particle $\sex$ is also an isomorph invariant, we can consider $X$ to be a function of $\sex$, i.e., $X=f(\Sex)$, and we have for the temperature derivative at constant density

\begin{equation}
\left(\frac{\partial X}{\partial \ln T}\right)_\rho = \frac{d X}{d\Sex} 
\left(\frac{\partial \Sex}{\partial\ln T}\right)_\rho =  \frac{dX}{d\Sex} c_V^{\textrm{ex}}, 
\end{equation}
where $c_V^{\textrm{ex}}$ is the configurational contribution to $c_V$, and is also isomorph invariant--that is, a unique function of $\Sex$. Thus the $T$-term in is also isomorph invariant. For the density derivative at constant temperature we use the identity 

\begin{equation}
\left( \frac{\partial\ln T}{\partial\ln\rho} \right)_X = - \frac{\left(\frac{\partial X}{\partial\ln\rho}\right)_T}
{\left(\frac{\partial X}{\partial\ln T}\right)_\rho}
\end{equation}
and note  (Eq.~\eqref{gamma_definition}) that the left hand side is just $\gamma$ (constant $X$ being the same as constant $\Sex$), thus

\begin{equation}
\left(\frac{\partial X}{\partial \ln \rho}\right)_T = - \gamma
\left(\frac{\partial X}{\partial \ln T}\right)_\rho
\end{equation}
Thus the density derivative is proportional to the (isomorph invariant) temperature derivative with a proportionality constant $-\gamma$. The latter is not isomorph invariant, but its variation over typically accessed densities
is small and can often be neglected (see Refs.~\onlinecite{Alba-Simionesco/Dalle-Ferrier/Tarjus:2013, Boehling/others:2012} for an exception). On the other hand the third term in  Eq.~\eqref{chi4NPTdecomp1} ($\rho$-term) includes also the factor ${\rho k_BT}/{K_T}$, which is the inverse reduced-unit isothermal bulk modulus. That it is not isomorph invariant follows from Eq.~\eqref{quasiuniversalU}, whose term $g(\rho)$ contributes non-invariant parts to volume derivatives of the (free) energy, such as the bulk modulus.

The conclusion from the above analysis is that $\chi_4^{\textrm{NPT}}$ is not isomorph invariant, while $\chi_4^{\textrm{NVT}}$ is. This applies also to the estimators introduced by Berthier {\it et al.} which involve dropping the NVE contribution:

\begin{align}
\chi_4^{\rm NVT} &\simeq \frac{1}{c_V/k_B}\left(\frac{\partial C(t)}{\partial \ln T}\right)_\rho^2 \quad {\rm (invariant)} \label{NVT_estimator}\\
\chi_4^{\rm NPT} &\simeq  \frac{1}{c_V/k_B}\left(\frac{\partial C(t)}{\partial \ln T}\right)_\rho^2 + \frac{\rho k_BT}{K_T}\left(\frac{\partial C(t)}{\partial \ln\rho}\right)_T^2 \quad {\rm (varies)} \label{NPT_estimator}
\end{align}
Thus isomorph theory predicts that there will be little variation along the isochrone when the $NVT$ version of $\chi_4$ is considered, so the latter is an allowed candidate for a quantity which controls the dynamics.

\section{Experimental tests}

\begin{figure}
\epsfig{file=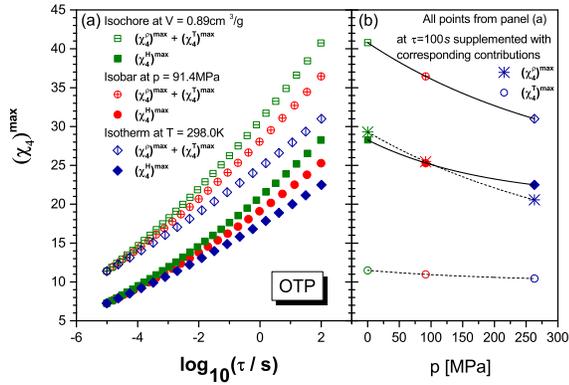,width=8 cm, clip=}
\caption{\label{OTP}
Experimental estimates of $\chi_4^{\rm NPT}$ for ortho-terphenyl taken from supplemental material to Ref.~\cite{Koperwas/others:2013} (Figure S.2) Panel (a) shows $\chi_4$ as a function of relaxation time along three different thermodynamic paths and estimated in two different ways: One (open symbols) is the estimate based on Eq.~\eqref{chi4NPTdecomp1}, namely Eq.~\eqref{NPT_estimator}, while the other is based on an alternative decomposition into contributions from
the NPH ensemble and a term involving the temperature derivative at constant
pressure \cite{Berthier/others:2007a}. Panel (b) shows the pressure dependence 
of the peak value on the isochrone
$\tau$=100s (top curve), together with the contributions 
induced separately by energy and density fluctuations ($T$- and $\rho$- terms, circles and stars respectively). The solid symbols indicate the estimate from the alternate decomposition. The lines are
guides for the eyes.
}
\end{figure}

Data for the NVT $\chi_4$ has already been published in Ref.~\onlinecite{Koperwas/others:2013}, in the supplementary material. We reproduce their figure S2 in Fig.~\ref{OTP}, which shows data from dielectric measurements made on the glass-forming liquid ortho-terphenyl over a range of temperatures and pressures.
In the left panel different estimates of $\chi_4$ are plotted as a function of relaxation time while in the right panel data for relaxation time 100s is plotted against pressure. Note that while it is in principle necessary to express the relaxation time in reduced units (rescaling by certain powers of temperature and density), at such viscous states the density and temperature change so little that the difference is negligible. In the right panel the lowest data-set (circles) gives the estimation of the $\chi_4^{\textrm{NVT}}$ from Eq.~\eqref{NVT_estimator}, which is fairly constant as pressure is varied along the isochrone, while the stars represent the third term in Eq.~\eqref{chi4NPTdecomp1} ($\rho$-term) which decreases as pressure (and density) increases, corresponding to the increase of the reduced bulk modulus. The top curve is the sum of the other two, representing the experimental estimate of $\chi_4^{\textrm{NPT}}$ via Eq.~\eqref{NPT_estimator}. It inherits the variation of the bulk 
modulus. Data for the latter (not shown) show a variation sufficient
to explain the variation here. The 
change of the estimate of $\chi_4^{\rm NPT}$ is
-24\%; the change of the estimate of $\chi_4^{\rm NVT}$ is -8\%. Note that in both
cases the term representing 
$\chi_4^{\rm NVE}$ is missing, so the true percentage change for $\chi_4^{\rm NPT}$ 
will be smaller. The 8\% change in the NVT estimate can be seen as a true 
deviation from isomorph invariance, since this quantity is formally isomorph 
invariant.

In contrast to Refs.~\cite{Grzybowski/others:2012a, 
Koperwas/others:2013,Grzybowski/others:2013}, Alba-Simionesco et al.  found an {\em increase} of the maximum of $\chi_4^{NPT}$ with increasing pressure for the liquid dibutyl-phthalate \cite{Alba-Simionesco/Dalle-Ferrier/Tarjus:2013}. In this case the
change in the reduced bulk modulus turns out to be relatively small, while there
is an unusually large change in $\gamma$ (denoted $x$ in that work), from 2.5
to 4, as density increases, thus the increase in $\chi_4^{\textrm{NPT}}$. 
The authors presented these results as being in contradiction to the RFOT theory, but the implication that there is a problem with the theory is not necessarily valid, given the use of a non-isomorph invariant definition of $\chi_4$.

\section{Simulations}

\begin{figure}
\epsfig{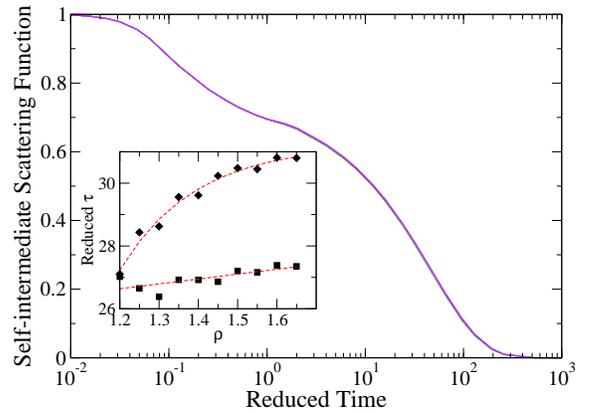}
\caption{\label{ISF_Isochrone}Self-intermediate scattering function for the Kob-Andersen system (large particles) for several state points along an isochrone with lowest density 1.20 (temperature 0.55), plotted against reduced time $\tilde t$. The arrow indicates increasing density. The inset shows the relaxation time $\tilde \tau$ defined as the time when the correlation function has fallen to $1/e$ along to curves: a isochrone (squares) determined by Eq.~\eqref{h_tilde_rho} parameterized by matching its logarithmic derivative to Eq.~\eqref{gamma_fluctuations} at density 1.6 and  and the configurational adiabat (diamonds) determined iteratively by fitting the logarithmic derivative of Eq.~\eqref{h_tilde_rho} to Eq.~\eqref{gamma_fluctuations} for each step in density of size 0.05. The isochrone and adiabat share the state point ($\rho=1.20,T=0.55$). Lines are to guide the eyes.}
\end{figure}

We have investigated the different contributions to $\chi_4$ in simulations.
The important results are that the NVT estimate is invariant along an  isochrone while the NPT estimator decreases, as expected. When the NVE 
contribution is included (which is not small), 
we find a slight {\em increase} in full NVT quantity. Thus
the true deviation from (otherwise expected)
isomorph invariance in that case has the opposite sign to what is seen in the 
NPT estimator.

Simulations were carried out on a binary Lennard-Jones system of 1024 particles using the Kob-Andersen 
potential parameters \cite{Kob/Andersen:1994} and usual composition of 80\% large particles; technical details are given in Appendix~\ref{technical_details}. A range of densities starting at the usual 1.2 and going up to 1.65 was simulated. Both NVE and NVT dynamics were used, but the thermostat relaxation time for the latter was chosen to be at each state point a few times the alpha relaxation time to give approximately constant energy during microscopic processes, to better correspond to the case of an ensemble of NVE trajectories as discussed in Sec.~\ref{isomorphInvarianceChi4}. In additional it was held constant in reduced units along isochrones.

Since an isomorph is an idealized concept which formally is both an isochrone and a (configurational) adiabat, while in practice isochrones and adiabats do not exactly coincide, we have considered both an isochrone and an adiabat.
To identify an isochrone we used the method of parameterizing the density scaling function $h(\rho)$ appearing in Eq.~\eqref{h_rho_T_const} that was used in Ref.~\onlinecite{Boehling/others:2012}. It is formulated in terms of a reference density $\rho_*$, with respect to which the relative density $\tilde\rho\equiv \rho/\rho_*$ is defined:

\begin{equation}\label{h_tilde_rho}
h(\tilde \rho) = \tilde \rho^4 (\gamma_*/2-1) - \rho^2 (\gamma_*/2 - 2)
\end{equation}
Here $\gamma_*$ is the scaling exponent at the reference density. For perfect isomorphs it should equal the fluctuation-based scaling exponent at $\rho_*$; in Ref.~\onlinecite{Boehling/others:2012} a value was chosen which gave a good collapse of the relaxation time data over a broad range of density and temperature: For reference density $\rho_*=1.6$, the value $\gamma_*=4.59$ was used. This can be used to generate a set of $\rho,T$ values which constitute an approximate isochrone, but it does not take account of the temperature dependence of $\gamma$. The value 4.59 was determined by a procedure which gives weight to more viscous state points; here we study less viscous state points due to the need for extra long runs for good $\chi_4$ statistics. The value of $\gamma$ observed from the fluctuations at density 1.6 is closer to 4.57; and we find indeed that we get a more exact isochrone if we generate the state points using this value. Note that this is a 0.5\% difference in $\gamma_*$; it is only relevant when trying to identify good isochrones on a linear (rather than logarithmic) time axis.

\begin{figure}
\epsfig{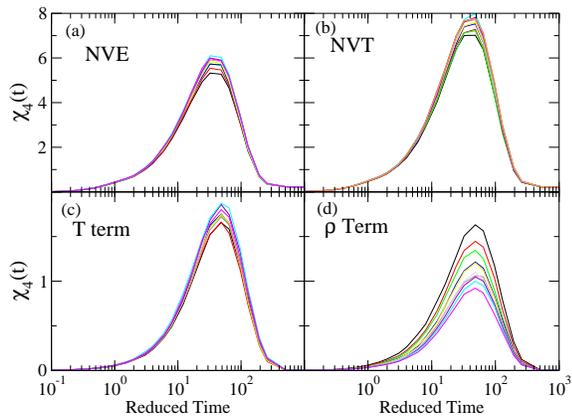}
\caption{\label{Chi4Collapse}(a) $\chi_4^{\textrm{NVE}}$ for the same points as in Fig.~\ref{ISF_Isochrone}; (b) $\chi_4^{\textrm{NVT}}$ for the same points; (c) Temperature term corresponding to experimental estimate of $\chi_4^{\textrm{NVT}}$ (d) Density term used to estimate $\chi_4^{\textrm{NPT}}$.
}
\end{figure}



Figure~\ref{ISF_Isochrone} shows data for an isochrone containing the point $\rho=1.2, T=0.55$. In the main panel the self-part of the intermediate scattering function is plotted for densities between 1.2 and 1.65 \footnote{Data are shown for the NVT ensemble; those for the NVE ensemble are indistinguishable in accordance with the Lebowitz formalism whereby the difference would involve $\langle\exp(i q \vec r(t))\rangle$ which is zero for an arbitrary tagged particle.}. The inset shows the relaxation time; it increases slightly, about 2\% over this range of densities. We have also simulated a configurational adiabat, identified as the curve whose logarithmic derivative $\left(\partial \ln T/\partial\ln \rho\right)_S$ is given by the fluctuation expression for $\gamma$ (Eq.~\eqref{gamma_fluctuations}) at each state point. The procedure used was the same as in Ref.~\onlinecite{Bailey/others:2013}: steps of 0.05 in density were taken, and Eq.~\eqref{h_tilde_rho} was re-parameterized at each step using the observed fluctuations (Eq.~\eqref{gamma_fluctuations}). For a system with perfect isomorphs these curves should be identical; for a real system they are close but not identical. A single parameterization of Eq.~\eqref{h_tilde_rho} using a relatively high reference density such as 1.6 (as opposed to re-parameterizing at each step in density) can generate a good isochrone over the whole simulated density range. But the observed $\gamma$ from fluctuations deviates from the that given by Eq.~\eqref{h_tilde_rho} as the density decreases (at density 1.20 the observed $\gamma$ is approximately 5.16 while the logarithmic derivative of $h(\tilde \rho)$, parameterized by $\gamma_*(\rho_{\rm ref}=1.60)=4.57$, is closer to 5.30). This indicates a deviation of the adiabat from the isochrone. It is intriguing that a single parameterization of  Eq.~\eqref{h_tilde_rho}--that is, determined by a single value of $\gamma_*$ at a given reference density, and used over the whole density range--generates a better isochrone than adiabat. On the latter the reduced relaxation time increases by about 10\% over the range of densities considered here (inset to Fig.~\ref{ISF_Isochrone}).

\begin{figure}
\epsfig{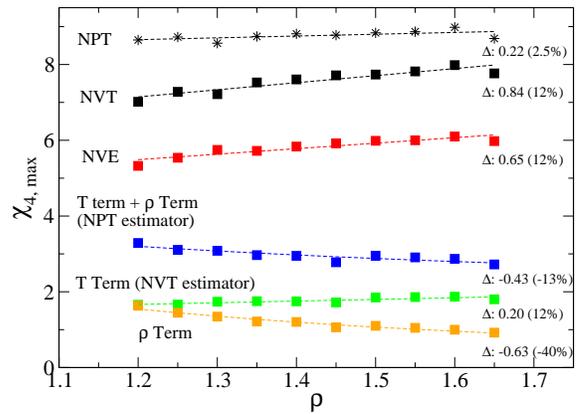}
\caption{\label{Chi4Maxima_I2}Maximum values of $\chi_4$ versus density for the isochrone in different ensembles also including different contributing terms. Absolute and percentage changes are indicated to the right of each set.}
\end{figure}

\begin{figure}
\epsfig{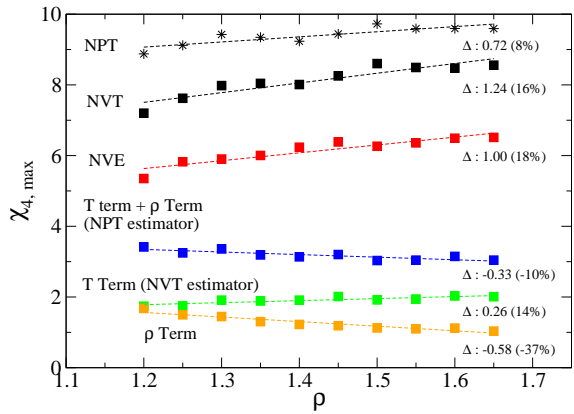}
\caption{\label{Chi4Maxima_Adiabat}Maximum values of $\chi_4$ versus density for the adiabat in different ensembles
also including different contributing terms.}
\end{figure}

Fig.~\ref{Chi4Collapse} shows $\chi_4$ for the approximate isochrone in the NVE and NVT ensembles, the T-term (or NVT-estimator) and the $\rho$-term (which in combination with the T-term gives the NPT-estimator). Reduced units for time are used. For the first three there is an approximate collapse, while for the $\rho$-term there is clearly no collapse.  This is in accordance with the expectation that the first three quantities are formally isomorph invariant while the fourth is not, while revealing that there is some variation even for the formally invariant quantities. A clearer idea of the trends as a function of density is given by considering the peak value of each $\chi_4(t)$ curve. These are shown in Fig.~\ref{Chi4Maxima_I2} for the isochrone and in Fig.~\ref{Chi4Maxima_Adiabat} for the adiabat. Also shown are the maxima of the true NPT quantity obtained by adding the $\rho$-Term to the NVT quantity. We have confidence that this is a good estimate; we have explicitly confirmed (data not shown) the decomposition of the NVT quantity, Eq.~\eqref{chi4_NVT_NVE_T_term}\footnote{Recall that by NVT quantity we mean with Newtonian dynamics, approximated in our simulations by a thermostat with a very long relaxation time; with a typical short relaxation time Eq.~\eqref{chi4_NVT_NVE_T_term} is not satisfied.}.

It can be seen that there appears, in fact, to be a slight increase in the NVE and NVT $\chi_4$. On the
other hand, the estimator of the NVT value by the temperature derivative is quite flat, as invariant as the correlation function itself, while the the estimator of the NPT value appears to decrease with increasing density, as expected from the behavior of the bulk modulus. Since the bulk modulus is not
formally isomorph invariant, this decrease is not surprising. The deviation seen in the NVE or full NVT quantity is interesting because it represents a true deviation from isomorph invariance in a formally isomorph invariant quantity. Unfortunately it does not appear in the experimentally accessible quantities. Ironically, when the full NPT quantity is considered, it is in practice perhaps the most invariant quantity due to a small percentage increase in the NVE and NVT quantities (a true deviation from isomorph invariance) and a large percentage decrease in the $\rho$-term (expected due to the bulk modulus). Since the latter term is relatively small the absolute changes tend to cancel, leaving the true NPT quantity quite flat.

\section{Discussion}

\subsection{Interpreting the extra contributions (T-term and $\rho$-term)}

One must be careful not to over-interpret the different effects of pressure (or density) and temperature on the dynamics. The dynamics can only said to depend on an isomorph invariant, for example the excess entropy $\Sex$. To show explicitly how this can be represented in the contributions to $\chi_4$, we re-write Eq.~\eqref{chi4NPTdecomp1} in terms of derivatives with respect to the alternative variables $\Sex$ and $\rho$ using the chain rule and standard thermodynamic identities:

\begin{align} 
\chi_4^{\textrm{NPT}} &- \chi_4^{\textrm{NVE}} = 
\frac{(c_V^{\textrm{ex}})^2}{c_V} \left(1 + \frac{\rho T \gamma^2 c_V}{K_T}  \right)
\left( \frac{\partial C}{\partial \sex}\right)_\rho^2 - \nonumber \\
&\frac{2\rho T \gamma c_V^{\textrm{ex}}}{K_T} \left(\frac{\partial C}{\partial\rho}\right)_{\sex} \left(\frac{\partial C}{\partial \sex}\right)_\rho +
\frac{\rho T}{K_T} \left(\frac{\partial C}{\partial \rho}\right)_{\sex}^2
\label{Lebowitz_Sex_P}
\end{align}

Note there is a cross term involving both derivatives in this expression. For a general choice of independent variables there will be such a term; the choice of temperature and density (or volume) is special because these variables are statistically independent as is known from thermodynamic fluctuation theory \cite{Reichl:1998}. We can interpret the coefficient of the first term by replacing $C$ with $\sex$ (since the above expression is just a Lebowitz-type formula \cite{Allen/Tildesley:1987,Lebowitz/Percus/Verlet:1967} with an extra factor $N$, we can certainly apply it to ordinary static quantities as well). Since the excess entropy is not a dynamical variable, its variance in the NVE ensemble is zero. Evaluating  Eq.~\eqref{Lebowitz_Sex_P} gives the coefficient 

\begin{equation}
\frac{(c_V^{\textrm{ex}})^2}{c_V} \left(1 + \frac{\rho T \gamma^2 c_V}{K_T} 
\right) \equiv \frac{(c_V^{\textrm{ex}})^2}{c_V}\left(1+R\right) = 
N \angleb{(\Delta \sex)^2}_{\textrm{NPT}}
\end{equation}
Thus this coefficient can be interpreted as the variance of excess entropy among the different $NVE$-members of the $NPT$ ensemble (note that it is not isomorph invariant due to the presence of $K_T$). The quantity 
$R\equiv \gamma^2 c_V /\tilde K_T=\gamma^2 c_v \rho k_BT/K_T$ was also defined by Dalle-Ferrier {\it et al.} \cite{Dalle-Ferrier/others:2007, Alba-Simionesco/Dalle-Ferrier/Tarjus:2013} to quantify the relative importance of density fluctuations compared to energy fluctuations on the dynamics. Here we see that more precisely it quantifies their relative contributions to fluctuations of excess entropy. Returning to $\chi_4$, in the case of an R liquid where $C(t)$ depends only on $\sex$, we get 

\begin{align}
\chi_4^{\textrm{NPT}} - \chi_4^{\textrm{NVE}} &= 
\frac{(c_V^{\textrm{ex}})^2}{c_V} \left(1 + R\right) \left(\frac{d C}{d \sex} \right)^2  \nonumber \\
&= N \angleb{(\Delta \sex)^2}_{\textrm{NPT}} \left(\frac{d C}{d \sex} \right)^2
\end{align}
Thus the variation of the dynamical quantity $\chi_4^{\rm{NPT}}$ along an isomorph is, for an R liquid, due to the thermodynamic fact that a contour of $\Sex$ is not a contour of the NPT-variance of $\Sex$.
Another way to interpret this is to consider the $NPT$-ensemble of $NVE$-trajectories at each point along an isomorph. Members whose energy and volume are the mean energy and volume experience dynamics isomorphic with corresponding members elsewhere on the isomorph. Other members have energy and volume near the mean values and will statistically be very similar to those on the mean. But the spread of volume  will vary along the isomorph: the ensemble will be narrower at high densities because the reduced bulk modulus typically increases with density. This is the sense in which the dynamical susceptibility is not invariant. Again we emphasize that the variation represented by the terms in a Lebowitz-type formula is variation of initial conditions, and not dynamical fluctuations.

\subsection{Ensemble-independent dynamic susceptibility}

\begin{figure}
\epsfig{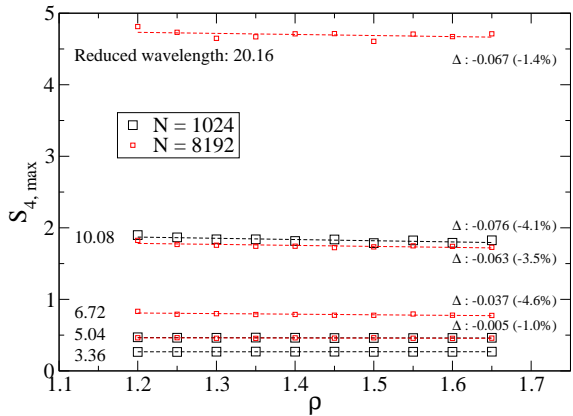}
\caption{\label{S4_maxima}  Maxima of $S_4(k, t)$ for different wavenumbers/wavelengths versus density on isochrone. The indicated wavelengths are in reduced units, and equal to $N^{1/3}/n$ for integer $n$. For wavelengths 10.08 and 20.16 a slight decrease of $S_{4,\rm{max}}$ is seen. Data from an system with double the linear size is also shown, for the exact same state points. Finite size effects cause the values of $S_4$ at wavelength 10.08 to differ slightly (the reduced relaxation time is 15\% shorter for the larger system, data not shown; its variation along the isochrone is as small as for the shorter one). For the largest $k$ the maximum value differs little from the long-time random value coming from the mean of $\cos^2$ (see Appendix~\ref{S4_definition}).}
\end{figure}

We consider now an issue raised by the argument of Flenner and Szamel that the $NPT$ version of $\chi_4$ is the correct one (for a pure substance; for a mixture extra terms accounting for concentration terms should be included) in the following sense \cite{ Flenner/Szamel:2010, Flenner/Szamel:2013}: Consider the four-correlation function $S_4(k, t)$ where $k$ should be distinguished from the wavenumber $q$ appearing in the self-intermediate scattering function. It denotes the wavevector associated with the Fourier transform of the correlator with respect to initial position (see Appendix~\ref{S4_definition} for details of our definition of $S_4$). At finite wavenumber $k$ the four-point correlation $S_4(k, t)$ or any wavenumber-dependent response function is ensemble-independent because the boundaries only couple to $k=0$ behavior--alternatively measuring at finite wavenumbers corresponds to sampling subsets of the system whose density, energy and concentration can fluctuate while their mean values are fixed. Taking the $k\rightarrow0$ limit {\em after the thermodynamic limit} gives therefore an ensemble-independent quantity. This raises the interesting question of the isomorph invariance of $S_4(k,t)$ at finite $k$. Its invariance in practice can be investigated in simulations, but what does the theory say--is $S_4(k, t)$ even formally isomorph invariant?  Answering this is not straightforward, but consideration of it suggests how the theory of isomorphs can predict its own limit of validity. In particular, a reconciliation of the non-invariance of the bulk modulus and the idea of an ensemble-independent definition of $\chi_4$ suggests that isomorph invariance may somehow break down at long length scales.

Consideration of the term $g(\rho)$ in the approximate potential energy Eq.~\eqref{quasiuniversalU}, which is responsible for the non-invariance of the bulk modulus, suggests how this might occur. This term is non-local, whereas the true potential energy is local. A way to restore locality to the approximation would be to replace $g(\rho)$ with a sum over particles $\sum_i g(\rho_{CG,i})$. Here $\rho_{CG,i}$ is a local density evaluated at the position of particle $i$, and involving a coarse-graining length $l_{CG}$. For a sufficiently large $l_{CG}$ this would make no difference to the hitherto documented isomorph invariant quantities, since these have all involved dynamics at short to moderate length scales. With such a local representation of potential energy for an R liquid we can predict that wavenumber-dependent response functions (at least those with a longitudinal component) such as $S_4(k, t)$ are formally isomorph invariant only for $k$ larger than a crossover value of order $1/l_{CG}$, while for smaller $k$ we can expect deviations.

This line of reasoning has two consequences. First, it turns the discussion about which ensemble is relevant into a one about which length scales are relevant. Second, the principle mentioned at the end of Section.~\ref{isomorphs} states that only isomorph invariant quantities can be relevant for determining isomorph-invariant measures of dynamics. This therefore suggests that density fluctuations and dynamical fluctuations at long wavelengths are not relevant for the usual measure of dynamics which involve relatively small wavelengths of order a few particle spacings. For a measure of dynamical heterogeneity satisfying the requirements of both ensemble independence and isomorph invariance one should consider $S_4(k,t)$ at some finite wavenumber $k$, which should be small enough to capture what correlated dynamics exists, but not arbitrarily small. This is tested in Fig.~\ref{S4_maxima}, which includes data for a system whose linear size is a factor of two larger ($N$=8192). Little variation is evident: a slight decrease at the largest two wavelengths. If we assume that the terms accounting for concentration fluctuations are at least as invariant as the T-term, then this is consistent with the fact that the $\chi_4^{\rm NTP}$ maximum also varies little (Fig.~\ref{Chi4Maxima_I2}). In that case the slope is positive, but the data are perhaps not precise enough to distinguish a change of +2.5\% from one of -1.4\%. Such little variation would normally be considered consistent with isomorph invariance, but the interpretation here becomes tricky: we already argued that the apparent invariance of the NPT quantity is due to cancellation of a true violation of isomorph invariance in the NVT quantity and the expected variation of the $\rho$-term. It is therefore not clear what can be concluded from these $S_4$ data. Investigation of a wider range of $k$ values (system sizes), temperatures (both more and less supercooled) and systems (including non-Roskilde liquids) is needed to resolve this issue.

\subsection{Open questions}

The analysis and discussion presented so far give rise to some open questions.

\begin{enumerate}
\item Is the cancellation of terms that leads to an almost invariant $\chi_4^{\rm NPT}$ an accident or does it represent something deeper going on? At more viscous state-points the T- and $\rho$- terms are expected to account for a larger fraction of $\chi_4^{\rm NPT}$, so the cancellation should be less effective. We see more or less the same behavior, however, at the lowest simulated temperatures (0.48 for the usual density 1.2; data not shown). As mentioned above, studies of less viscous state points and non-R systems are needed to clarify the general picture.
\item The line of reasoning about long-wavelength fluctuations suggests that the static structure factor $S(q)$ must show deviations from isomorph invariance at sufficiently small $q$ (this is also required by the non-invariance of the bulk modulus). This should be investigated using large systems. Note that no deviations have been found in the (coherent) intermediate scattering function at small $q$: Veldhorst et al studied the intermediate scattering function in a polymer model and found invariance at all wavenumbers \cite{Veldhorst/Dyre/Schroder:2014}. Extensive testing of this has yet to be done, however.
\item The definition of $S_4$ employed here (see appendix \ref{S4_definition}) includes contributions where the two wave-vectors ($\bfa{k}$ and $\bfa{q}$) are parallel with each and ones where they are perpendicular. A decomposition into terms where the vectors $\bfa{k}$ and $\bfa{q}$ are parallel versus perpendicular, as done by Flenner et al. \cite{Flenner/Staley/Szamel:2014}, might show interesting differences regarding isomorph invariance. In particular one could speculate, by analogy with the bulk versus shear modulus, that the longitudinal (parallel) case might show greater deviations from isomorph invariance  than the transverse (perpendicular) case. Similarly the wavenumber-dependent bulk and shear viscosities would be worth investigating at small $q$. Further theoretical work is required to elucidate the question of which wavevector-dependent quantities, if any, are expected to be invariant at low $k$. 
\item Another formally invariant quantity that increases slightly along an isomorph (adiabat) for Lennard-Jones liquids is $C_V$ \cite{Bailey/others:2013}. In Ref.~\onlinecite{Bailey/others:2013} it was shown that whether $C_V$ increases or decreases depends on a certain feature of the pair potential and one might speculate that $\chi_4^{\rm NVT}$'s behavior depends similarly on the potential. While we have no theoretical argument for this, it is straightforward to check empirically.
\end{enumerate}

\section{Conclusion}

We have analyzed the different terms in the decomposition 
\eqref{chi4NPTdecomp1} which is the basis of experimental estimations of
 $\chi_4$ in pure substances, with a view to determining their formal isomorph
invariance. We find that the $\rho$-term, which accounts for volume fluctuations, spoils isomorph invariance due to its containing the bulk modulus, which is known to not be formally isomorph invariant. The fact that $\chi_4^{\rm NPT}$ is not formally invariant has implications both for isomorph theory and for the physics of glass forming liquids: Isomorph invariance cannot be expected to hold for dynamics at arbitrarily long wavelengths (take the bulk sound
velocity, for example) while the usual measures of liquid dynamics can therefore not depend, or be controlled by, measures of long-wavelength fluctuations 
in the dynamics. 

Data from simulations confirm the non-invariance of the $\rho$-term, but also provide a means to check the NVE-contribution which is not usually accessible experimentally. Interestingly, it too shows a variation, which tends to cancel that due to the $\rho$-term. Thus the true NPT quantity turns out actually to be quite invariant, although theory does not predict it to be so.

\acknowledgments

The center for viscous liquid dynamics ``Glass and Time'' is sponsored by the Danish National Research Foundation's grant DNRF61. The authors are grateful for comments on a previous version of this manuscript by G. Szamel and L. Berthier.

\appendix

\section{\label{technical_details}Technical details of simulations}

\begin{figure}
\epsfig{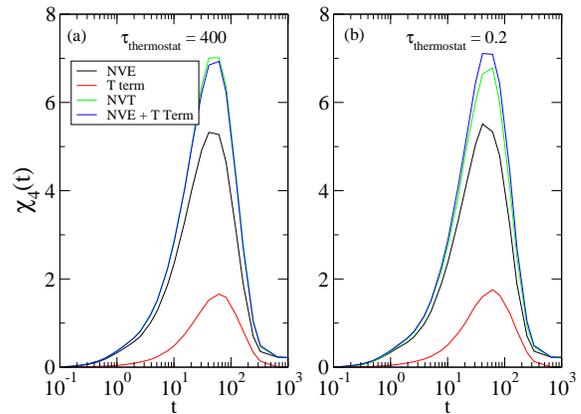}
\caption{\label{LongShortTau}Effect of varying the thermostat relaxation time, $\rho=1.20$, $T=0.55$ ($\tau_\alpha=27$). For a thermostat relaxation time long compared to $\tau_\alpha$, and a simulation long enough to include of order 10000 thermostat relaxation times, we have effectively Newtonian (energy-conserving) dynamics with an NVT ensemble. In this case Eq.~\eqref{chi4_NVT_NVE_T_term} holds, as can be seen (a). For the case of a very short thermostat relaxation time, there is a small but significant difference (b).}
\end{figure}

A ``shifted forces'' 
cutoff of 2.5$\sigma$ was used \cite{Toxvaerd/Dyre:2011}.
The software was RUMD \cite{RUMD:2014} running on nVidia 
graphical processing units. The time step was 
chosen such that its reduced value was same for all points along the isomorph; 
for $\rho=1.2, T=0.55$ the value was 0.005 in ordinary MD units (based on the 
Lennard-Jones length and energy parameters for the large particles). 
For the runs where $\chi_4$ was calculated directly, the simulation run length 
corresponded to at least $10^4$
relaxation times. For the runs at neighboring densities and temperatures, in
order to numerically differentiate the correlation function, it was $10^3$ 
relaxation times. 
The $q$-value was chosen equal to that used by Kob and Andersen (7.25
for large particles) at density 1.20, but scaled proportional to $\rho^{1/3}$ for
other densities: $q_\rho = q_{1.2}(\rho/1.2)^{1/3}$. Numerical 
differentiation was carried out at fixed $q$ and $t$ (i.e., in ordinary rather 
than reduced units). The relaxation time for the thermostat was chosen to be fixed in reduced units,
and as mentioned above, at least a few times the relaxation time. Under these circumstances the energy is effectively constant on the time scale of relaxation processes, and the premise of Eq.~\eqref{chi4_NVT_NVE_T_term}, namely a canonical distribution of fixed-energy trajectories, is realized. See Fig.~\ref{LongShortTau} for an illustration of the effect of thermostat relaxation time.

\section{\label{S4_definition}Definitions of 2- and 4-point correlation functions}

To give a precise definition of the two-point and four point correlators we use, we start by defining a single particle, two-time quantity $f_{q,i}(t)$:

\begin{equation}
f_{q,i}(t) \equiv \frac{1}{3}\left( \cos(q \Delta x_i(t)) +
\cos(q \Delta y_i(t)) + \cos(q \Delta z_i(t))  ) \right)
\end{equation}
Contributions from all three coordinates are included to give extra averaging.
We define the summed correlator $F^A(q, t)$ as the average over type A particles:

\begin{equation}
F^A(q,t) = \frac{1}{N_A}\sum_i^{N_A} f_{q,i}(t).
\end{equation}
The expectation value of this gives the self-intermediate scattering function:

\begin{equation}
F_s(q,t) = \angleb{F^A(q, t)} = \frac{1}{N_A}\angleb{\sum_i^{N_A} f_{q,i}(t)}
\end{equation}
The variance of the correlator $F^A$, multiplied by $N$ (the total number of particles) is the dynamic susceptibility, as described in the main text. We note that at long times when particle positions have de-correlated from their initial values the variance is not zero but $\frac{N}{N_A} \frac{3}{9}\angleb{\cos^2}$ which is 0.21 for the composition $N_A/N=0.8$. This can be seen in the Fig.~\ref{Chi4Collapse} (a) and (b), and differs from the case where the dynamical correlator is defined using an overlap function instead of a cosine.

In order to define the four-point function $S_4$ we first define a quantity $\rho(k,q,t)$ as the Fourier transform with respect to initial positions ($x$-coordinate) of correlator,

\begin{equation}
\rho(k, q, t) \equiv   \frac{1}{N_A} \sum_i^{N_A}  f_{q,i}(t) \exp(i k x_i(0)),
\end{equation}
(note that for $t\rightarrow0$, any $q$ this is just Fourier mode $k$ of the density of A particles). Finally the four point correlation function $S_4(k,t)$ (suppressing the $q$-dependence for more convenient notation) as

\begin{equation}
S_4(k, t) \equiv N \angleb{\rho(k,q,t) \rho(-k,q,t)} - (F_s(q,t))^2 \delta_{k,0}
\end{equation}
Note that $N$ is the total number of particles of all types. Setting $k=0$ in this definition gives (the ensemble-dependent) $\chi_4$ as defined in the text; taking the $k\rightarrow 0$ limit after the thermodynamic limit gives an ensemble-independent quantity.


\end{document}